\documentclass[twocolumn, article, secnumarabic ,amssymb, amsmath, nobibnotes, prx, 10pt, groupedaddress, superscriptaddress, notitlepage]{revtex4-2}

\usepackage[usenames,dvipsnames]{xcolor}

\usepackage[utf8]{inputenc}
\usepackage{float}
\usepackage{graphicx}
\usepackage{bm}
\usepackage{mhchem}
\usepackage{soul}
\begin{document}

\title{Controlling Solvent Quality by Time: \\ Self-Avoiding Sprints in Nonequilibrium Polymerization}

\author{Michael Bley}%
\email[Michael Bley: ]{michael.bley@physik.uni-freiburg.de}
\affiliation{Applied Theoretical Physics – Computational Physics, Physikalisches Institut, Albert-Ludwigs-Universität Freiburg, Hermann-Herder Strasse 3, D-79104 Freiburg,Germany}

\author{Upayan Baul}
\affiliation{Applied Theoretical Physics – Computational Physics, Physikalisches Institut, Albert-Ludwigs-Universität Freiburg, Hermann-Herder Strasse 3, D-79104 Freiburg,Germany}
\author{Joachim Dzubiella}
\email[Joachim Dzubiella: ]{joachim.dzubiella@physik.uni-freiburg.de}
\affiliation{Applied Theoretical Physics – Computational Physics, Physikalisches Institut, Albert-Ludwigs-Universität Freiburg, Hermann-Herder Strasse 3, D-79104 Freiburg,Germany}
\affiliation{Cluster of Excellence livMats@FIT - Freiburg Center for Interactive Materials and Bioinspired Technologies, Albert-Ludwigs-Universität Freiburg, Georges-Köhler-Allee 105, D-79110 Freiburg, Germany}

\date{\today}%
\begin{abstract}
A fundamental paradigm in polymer physics is that macromolecular conformations in equilibrium can be described by universal scaling laws, being key for structure, dynamics, and function of soft (biological) matter and in the materials sciences. Here, we reveal that during diffusion-influenced, {\it nonequilibrium} chain-growth polymerization, scaling laws change qualitatively, in particular, the growing polymers exhibit a surprising self-avoiding walk (SAW) behavior in poor and $\theta$-solvents. Our analysis, based on monomer-resolved reaction-diffusion computer simulations, demonstrates that this phenomenon is a result of i) nonequilibrium monomer density depletion correlations around the active polymerization site, leading to a locally directed and self-avoiding growth, in conjunction with ii) chain (Rouse) relaxation times larger than the competing polymerization reaction time. These intrinsic nonequilibrium mechanisms are facilitated by fast and persistent reaction-driven diffusion ("sprints") of the active site, with analogies to pseudo-chemotactic active Brownian particles. Our findings have implications for time-controlled structure formation in polymer processing, as in, e.g., reactive self-assembly, photo-crosslinking, and 3D printing.
\end{abstract}
\maketitle

\section{Introduction}

Polymer physics provides universal scaling concepts~\cite{Flory1953,  Lifshitz1978, DeGennes1979, Doi1986, Rubinstein2003} which have strongly shaped the research in the natural and life sciences as well as engineering in the last decades~\cite{Chan1990, Onuchic1997, Pande2000, Ballauff2004, Hofmann2012, Brangwynne2015, Muelhaupt2017}.  Paradigmatic are simple scaling concepts in equilibrium, e.g., for the end-to-end distance of a chain-like polymer, $R_{ee} \propto N^{\nu}$, where the conformational scaling exponent $\nu$ depends on the solvent quality. For a purely random walk (ideal or $\theta$-solvent), $\nu\simeq 1/2$, while for poor solvents the polymer collapses and $1/3 < \nu \lesssim 1/2$, and for athermal solvents $\nu  \simeq 3/5$, constituting the well-known self-avoiding walk (SAW) or 'Flory' scaling~\cite{Flory1953} for swollen conformations. Most polymers exhibit a universal scaling behavior of various structural and dynamic properties, such as coil dimensions and relaxation, despite a different chemical composition.  This universality (in equilibrium) across length and time scales formed the basis for the extraordinary historical success of polymer physics in a wide range of fields. 

However, functional materials are typically synthesized, processed, and functioning under {\it nonequilibrium} conditions. Hence, growing attention has been drawn to investigate nonequilibrium polymer properties, how to conserve them,  and their consequences on material design. In particular, the possibility of generating stored, 'extra' free energy, stresses, and memory emerges, which can be potentially harvested for the design of highly responsive, interactive, or even adaptive materials~\cite{Stuart2010,Thomas2011, Liu2018, Walther2019}. The anticipated wide range of new structural, dynamical, and mechanical properties arises from the nonequilibrium competition between the time scales of (reaction) synthesis, processing, polymer relaxation, and observation (function)~\cite{DeGennes1982a, Toan2008, Alexander-Katz2009, Chandran2019, Reiter2020, Guerin2012, Katkar2018, Chubak2020}. However, such a complex competition of time scales, often also involving spatial modulation (i.e., 'spatiotemporal correlations'), has hampered our understanding of nonequilibrium polymer properties up to date. 

For synthesizing strongly 'out-of-equilibrium polymers', the type of kinetics and in particular the speed of the polymerization reaction play a key role. Most of the common and commercially used polymer architectures are produced by chain polymerization techniques~\cite{Flory1953, Penczek2008} such as radical polymerization. Controlling (fast) photo-polymerization reactions is crucial for mastering high-resolution 3D and 4D printing techniques~\cite{Muelhaupt2017, Bagheri2019, Telitel2020}. Polymerization is also possible on colloidal scales by guided, diffusion-controlled self-assembly of supracolloidal polymer chains~\cite{Sciortino2007, Lu2008, Groeschel2013}. However, structural insights are sparse and quantitative kinetic rate laws still absent.  Importantly, experiments observe large macroscopic changes during and after synthesis, e.g., retarded material shrinkages associated with the fast polymerization during the autoacceleration phase~\cite{Tran-Cong-Miyata2017} (also Trommsdorff-Norrish or gel effect~\cite{Schulz1956, Achilias1992}), whose microscopic mechanisms remain unexplained. 

On the fundamental level, magnetic-tweezer experiments of the real-time dynamics of a growing polymer indeed demonstrated a competition between reaction and conformational relaxation times~\cite{Liu2017}. Studies on chain walking catalysis of dendritic polymers \cite{Dockhorn2019} also reported a dependence of the emerging structure on the reaction rate: High rates lead to linear structures, while lower ones formed hyperbranched structures, hinting to a nonequilibrium conformational behavior tunable by the rate. While molecular computer simulations of growing polymer chains have been performed~\cite{Akkermans1998, Perez2008, Farah2012, DeBuyl2015, Liu2016}, they either assumed good solvent conditions or did not systematically investigate nonequilibrium behavior in varying solvent qualities.  However, classical theoretical concepts for diffusion-influenced polymerization kinetics~\cite{Schulz1956, DeGennes1982a, Russell1988, Achilias1992} shall strongly benefit from studies with a spatiotemporal, particle-level resolution including memory or proximity effects.  

Here, using monomer-resolved reaction-diffusion computer simulations of chain-growth polymerization of a single polymer, we demonstrate that solvent quality can be controlled by time scales in nonequilibrium, and also depends on the time of of observation.  In other words, we observe substantial nonequilibrium effects on the conformational properties of the growing chain: for fast polymerization, the time-dependent size of the polymer exhibits an unexpected SAW scaling in ideal and $\theta$-solvent conditions, and also enhanced scaling exponents in poor solvents, before they relax back into equilibrium. Our surprising findings can be explained by intrinsic nonequilibrium spatiotemporal correlations leading to local structural depletion and directed growth of monomers, related to pseudo-chemotactic active Brownian particles (ABPs) \cite{Bechinger2016} steered by nuitrition concentration gradients~\cite{Lapidus1980, Merlitz2020}.  The key for understanding is the competition of the various important time scales (diffusion, reaction, chain relaxation, observation) in the system, realizable only in a certain parameter window of intrinsic reaction propensities and monomer densities,  which we discuss in detail. Our results will be useful for various materials applications where structure formation in nonequilibrium synthesis and processing can be controlled by time.  

\begin{figure*}[!htbp]
\begin{center}
\includegraphics[width=17.8cm]{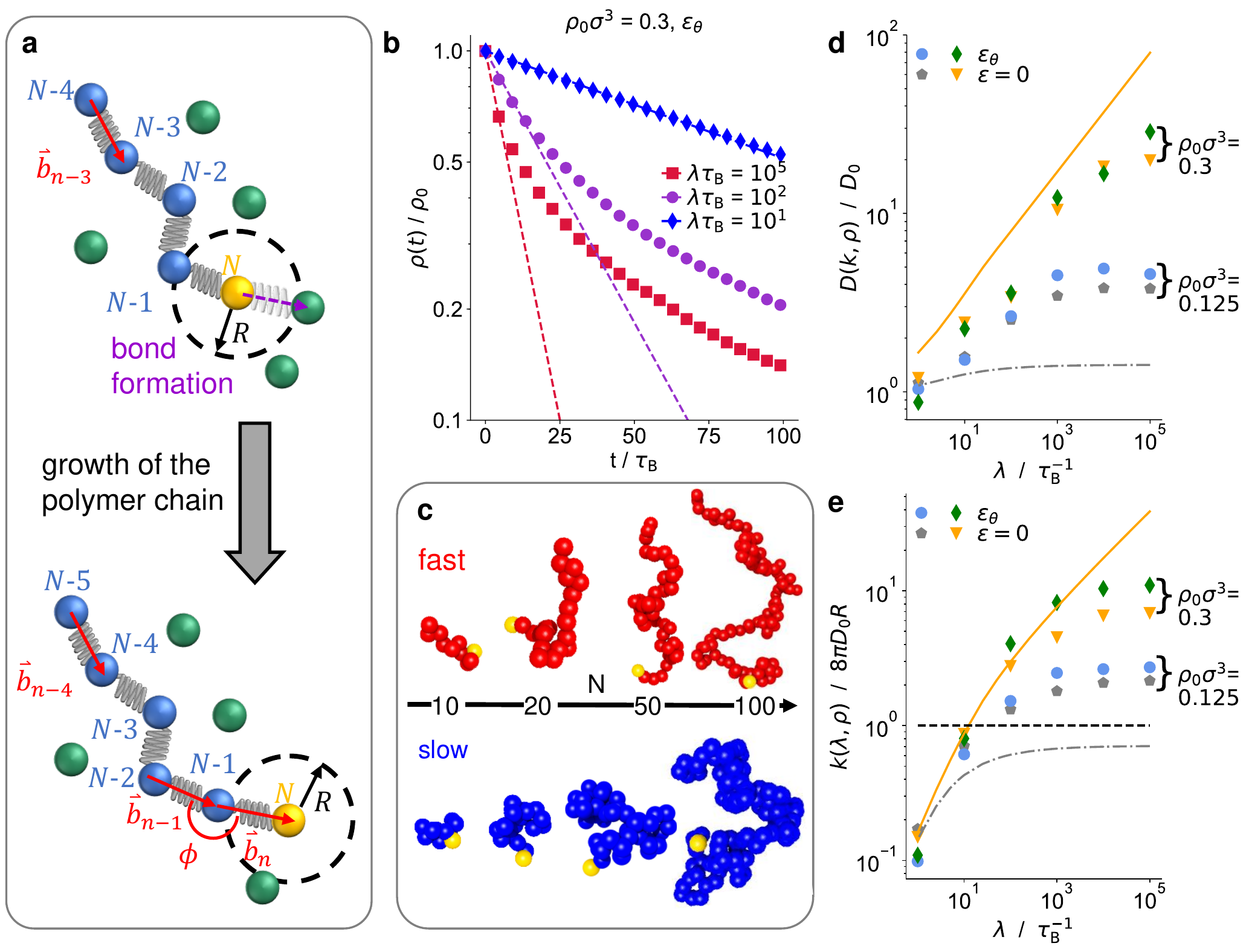}
\caption{\textbf{Simulation and kinetics of a growing polymer chain.} (a) Schematic representation of the growth step of a polymer chain (blue) with an active center (AC, yellow) with reactive radius $R$ and reaction propensity $\lambda$. During a reaction a bond is formed between a free monomer of size $\sigma$ (green) and the AC to propagate the chain. The newly added monomer spans up an angle $\phi$ by the terminal two bond vectors $\vec{b}_{n-1}$ and $\vec{b}_{n}$ during bond formation. The newly attached monomer becomes the new active center. (b) Concentration of free monomers $\rho(t)/\rho_{0}$ scaled by its initial value $\rho_{0}\sigma^{3} = 0.3$ for fast (red; $\lambda = 10^{5}$ $\tau_{\mathrm{B}}^{-1}$), intermediate (magenta; $\label = 10^{2}$ $\tau_{\mathrm{B}}^{-1}$), and slow (blue; $\lambda = 10^{1}$ $\tau_{\mathrm{B}}^{-1}$) reactions versus time $t$ (scaled by the Brownian time scale $\tau_B=\sigma^2/D_0$) at $\theta$-conditions, $\varepsilon = \varepsilon_{\theta}$. Symbols represent simulated data, the lines show the corresponding fits to initial times according to the first order rate equation (\ref{RateLaw:eq}). (c) Snapshots of fast (red) and slowly (blue) growing chains. (d) Total diffusion coefficient $D(\lambda, \rho)$ versus reaction propensity $\lambda$ for perfectly ideal ($\varepsilon=0$) and real chain $\theta$-conditions, $\varepsilon=\varepsilon_\theta$ (simulation: symbols; dash-dotted gray and orange line: numerical solutions from the set of coupled Eqs. (\ref{Rate:eq1}) and (\ref{DiffFast:eq}), respectively). (e) Scaled reaction rate $k(\lambda, \rho)$ for the same parameters (simulation: symbols; dash-dotted gray and orange line: numerical solutions from the set of coupled Eqs. (\ref{Rate:eq1}) and (\ref{DiffFast:eq}), respectively); horizontal dashed black line: fastest Smoluchowski bimolecular reaction rate $k_{\mathrm{S}} = 8 \pi D_{\mathrm{0}}R$.}
\label{Fig1:fig} 
\end{center}
\end{figure*}
\noindent

\section{Reaction rates of growing chains}
\subsection{Macroscopic rate laws}
In our reactive Brownian dynamics simulation framework, the chain polymerization propagates as
\begin{equation}
\ce{P_{\textit{N}-1}-AC + M} \xrightarrow{k(\lambda)} \ce{P_{\textit{N}}-AC} \; \; \; \; \; \mathrm{with} \; N = 1 ... N_{M,0} \;, 
\label{Propagation:eq}
\end{equation}
where $\ce{P_{\textit{N}-1}-AC}$ is a polymer chain consisting of $N-1$ non-reactive monomers and with a terminal active center (AC), which reacts with a free monomer M in the surrounding solution with monomer density $\rho(t)$ irreversibly with a reaction rate constant $k(\lambda)$ to form a chain with one additional bonded monomer $\ce{P_{N}-AC}$. Fig.~\ref{Fig1:fig}a illustrates a propagation step: if free monomers are found within the reactive volume of the AC, then a new bond is formed between the AC and the closest free monomer with a reaction frequency or "propensity" $\lambda$. The latter interpolates between reaction-control and diffusion-control~\cite{Achilias1992,Dibak2019}. The quality of the (implicit) solvent in our simulations is tuned by the interaction parameter $\varepsilon$ in the Lennard-Jones interaction between the monomers: $\varepsilon=0$ is a perfectly ideal chain and $\varepsilon_\theta$ describes real-chain $\theta$-conditions, such that $\varepsilon<\varepsilon_\theta$ are good solvent conditions and  $\varepsilon>\varepsilon_\theta$ are poor solvent conditions in equilibrium (see Appendix \ref{SIsec1:sec}). 

Typically such reaction kinetics would be described as pseudo-first order~\cite{Atkins2010}, since the concentration of the AC, $\rho_{\mathrm{ac}} = V^{-1}$, remains constant in the total volume $V$. The concentration of free monomers is initially $\rho_{0} =\rho(t=0)= N_{\mathrm{m},0}/V$ with $N_{\mathrm{m},0}$ as the initial monomer concentration. The elementary rate law and its integrated form for the time-dependent concentration $\rho(t)$ of free momomers then write
\begin{equation}
\label{RateLaw:eq}
\frac{\mathrm{d} \rho(t)}{\mathrm{d}t} = - k(\lambda)\rho(t) \, \rho_{\mathrm{ac}} \; \; \; \; \; \mathrm{and} \; \; \; \; \; \rho(t) = \rho_{0}  e^{- t k(\lambda) \rho_{\mathrm{ac}}} \; , 
\end{equation}
where the degree of polymerization is $N(t) = -V{\rm d}\rho(t)/{\rm d}t$. This simple first-order kinetics says that the concentration of monomers should decay exponentially with a time scale set by $V/k(\lambda)$ and the polymer grows with a time scale (inverse rate) $\tau_{\rm react}=(k(\lambda)\rho(t))^{-1}$. 

Reactions are fast (i.e., diffusion-influenced or controlled) if the propensity $\lambda$ is large and/or the free monomer density $\rho$ is large. Simulation results for the polymerization are exemplified in Fig.~\ref{Fig1:fig}b for three selected reaction propensities $\lambda$ and initial monomer bulk density $\rho_{0}\sigma^{3} = 0.3$, ($\sigma$ is the size of a monomer  and sets our length scale). The snapshots taken from growing chains at high and low $\lambda$ (red and blue in Fig.~\ref{Fig1:fig}c) already indicate that they exhibit differences in the topological growth behavior: A fast growth leads to more extended chains, whereas slower reaction propensities yield a more compact, sometimes even globular chain structure. Importantly, in Fig.~\ref{Fig1:fig}b we find that the elementary law~(\ref{RateLaw:eq}), fitted to the data for initial times cannot even qualitatively describe the overall polymerization rate for the faster reactions, that is, at the higher propensities clear deviations are observed for increasing polymerization time.  \\
\subsection{Microscopic description of reaction rate coefficients}
One dominant reason for the failure of first-order kinetics is the coupling between the reaction itself and a reaction-driven jump-diffusion of the active end-monomer~\cite{Russell1988,Achilias1992}, which was introduced already in the seminal work by Schulz~\cite{Schulz1956} as 'reaction-diffusion' and verified experimentally~\cite{Anseth1994}. Theories for diffusion-influenced bimolecular reactions require these diffusion coefficients, e.g., as in the here relevant Doi-scheme~\cite{Doi1975a, Dibak2019}
\begin{equation}
k(\lambda) = 4 \pi D R \left[ 1 - \frac{\tanh\left( R \sqrt{\frac{\lambda}{D}}\right)}{R \sqrt{\frac{\lambda}{D}}} \right] \; ,
\label{Rate:eq1}
\end{equation}
where $R = \sqrt[6]{2} \sigma$ represents the reaction radius around the active center. The Doi rate approaches the classical Smoluchowski rate \cite{Achilias1992, Dibak2019} for complete diffusion-control, $k = 4 \pi D R$, for infinitely fast propensities, $\lambda \rightarrow \infty$.
\subsection{Coupling between active center diffusion and reaction rates} 
The mutual diffusion coefficient $D$ between particles A and B is $D = D_{\mathrm{A}} + D_{\mathrm{B}}$. In the case of a growing polymer chain, thus~\cite{Achilias1992}
\begin{equation}
D(k, \rho) = D_{\mathrm{m}} + D_{\mathrm{ac}}(k, \rho) = D_{\mathrm{m}} + D_{\mathrm{tb}} + D_{\mathrm{jump}}(k, \rho) \; , 
\label{DiffFast:eq}
\end{equation} 
consisting of the contribution by the free monomers $D_{\mathrm{m}}$, and the reaction-rate dependent diffusion of the chain's active center $D_{\mathrm{ac}}(k, \rho)$. The latter is the sum of the thermal, fluctuation-induced diffusion of the terminal bead $D_{\mathrm{tb}}$ and a jump (reaction-)diffusion $D_{\mathrm{jump}}(k, \rho)$. $D_{\mathrm{tb}}$ depends on the relaxation time scales of a polymer chain consisting of $N$ monomers, which are described by the Rouse model \cite{Rouse1953, Doi1986, Rubinstein2003}: A chain segment of length $N/p =: s$ relaxes with a time scale $\tau_{p} \approx b^{2}/(6\pi^{2}D_{\mathrm{m}}) s^{2}$. The Rouse time $\tau_{\rm Rouse}$ is the (longest) relaxation time of the whole chain ($s = N$), while $\tau_{\mathrm{m}} \approx b^{2}/(6\pi^{2}D_{\mathrm{m}})$ represents the Kuhn monomer relaxation time for the highest mode, $s=1$. Let us compare the reaction time scales $\tau_{\mathrm{react}} = (k \rho)^{-1}$ with $\tau_{\mathrm{m}}$ and $\tau_{\mathrm{Rouse}}$. For short times $\tau_{\mathrm{react}} \lesssim \tau_{\mathrm{m}}$, the fastest diffusive behavior with $D_{\mathrm{tb}} \simeq D_{\mathrm{m}}$ should apply. At long reaction time scales ($\tau_{\mathrm{react}} \gg \tau_{\mathrm{Rouse}}$), the slowest diffusive behavior following the chain center-of-mass with $D_{\mathrm{tb}} \approx D_{\mathrm{m}}/N$ is recovered~\cite{Rouse1953, Rubinstein2003}. Hence, depending on the reaction time scale, different diffusion scales are relevant for the AC motion and the growth process. 

The second contribution, $D_{\mathrm{jump}}(k, \rho) = \alpha b^{2} \rho k(\lambda)/6$, covers a 'jump' process, since after the formation of a new bond in the chain, the AC moves discretely by one bond length $b$ to be the new terminal chain element. This maps to diffusive, random-walk behavior in 3D with step length $b$ and jump rate $\rho k(\lambda)$~\cite{Schulz1956, DeGennes1982a, Russell1988, Achilias1992}. Importantly, the jump diffusion is determined by the reaction rate itself, thus requiring a self-consistent solution of Eqs.~(\ref{Rate:eq1}) and (\ref{DiffFast:eq}) which we explicitly perform below. Moreover, in this work we uncover that fast growth leads to a locally directed, less random motion of the AC. We consider this effect by introducing the coefficient $\alpha = \tau_{\mathrm{rot}}/\tau_{\mathrm{react}} = \tau_{\mathrm{rot}}  \rho k(\lambda)$. The rotational reorientation time $\tau_{\mathrm{rot}}$ describes in analogy to Active Brownian Particles (APBs)~\cite{Bechinger2016} the direction memory, or persistence, of translational motion. Here, the lower limit $\alpha = 1.0$, where $\tau_{\mathrm{rot}}=\tau_{\mathrm{react}}$ (that is, completely random attachment of a new monomer) represents no persistence, while $\alpha > 1.0$ and thus $\tau_{\mathrm{rot}} > \tau_{\mathrm{react}}$ covers a directed growth behavior. The latter leads to enhanced long-time diffusion as for ABPs~\cite{Bechinger2016}, since between multiple growth events the chain does not loose its directional memory.\\

These considerations lead to lower and upper limits for the mutual diffusion $D(\lambda, \rho)$ and the rate $k(\lambda)$ to describe our simulation data. The slow limit is obtained by assuming the longest Rouse relaxation and $\alpha=1$, involving no fit parameter. The fast limit assumes the fastest, short-time diffusion for the terminal bead and uses $\alpha$ as a fit parameter to describe our fastest data. For $\rho_{0}\sigma^{3} =0.3$, a value of $\alpha = 1.41$ is found. Fig.~\ref{Fig1:fig}d displays the mutual diffusion constant $D(\lambda, \rho_{0})$ from simulations (symbols) compared with the self-consistent numerical solution of Eqs.~(\ref{Rate:eq1}) and (\ref{DiffFast:eq}) for the fast and slow diffusion limits, where we used $\rho = \rho_{0}$ for simplicity in our qualitative discussion. In Fig.~\ref{Fig1:fig}e we show a comparison of the corresponding rates. Note that  the simulated rate can be one order of magnitude faster than the classical Smoluchowski limit, because the jump diffusion leads to fast "sprints" of the AC. The figure thus shows that for propensities $\lambda > 10~\tau_{\mathrm{B}}^{-1}$, not only the diffusion contribution of the terminal bead $D_{\mathrm{tb}}$, but also the inclusion of $D_{\mathrm{jump}}$ is required to describe the simulated reaction rate coefficients (symbols). Our analysis demonstrates that our improved diffusion rate theory serves as a very reasonable orientation for the accessible range of diffusion coefficients for growing polymer chains, while quantitative rate descriptions are still very challenging due to the complex competition of time scales, and nonequilibrium structural effects as we will explore deeper below.  

We note that in the diffusion-controlled limit $\lambda \rightarrow \infty$ (and assuming $\alpha=1$ for simplicity) we derived an analytical expression for the total (self-consistent), yielding
\begin{equation}
k_{\mathrm{dc}}(\rho) = \frac{4 \pi (D_{\mathrm{m}} + D_{\mathrm{tb}})R}{1 - \frac{2}{3} \pi R b^{2} \rho} \, .
\label{Rate:perc}
\end{equation}
The result is remarkable for two reasons: i) during the reaction $\rho(t)$ decreases and the rate slows down beyond first order kinetics, which explains our observations in Fig.~\ref{Fig1:fig}b). And, ii), the denominator of equation (\ref{Rate:perc}) diverges at a critical monomer density of $\rho_{\mathrm{crit}} = (2/3 \pi R b^{2})^{-1} \approx 0.43 \; \sigma^{-3}$. The physics behind this behavior is that for higher densities of the free monomers, the jump diffusion of the active center becomes so fast that after every jump a new reacting monomer will be found with high probability. Very fast cascade reactions (very fast "sprints") become possible where segments grow almost instantaneously and the rate diverges. The critical density $\rho_{\mathrm{crit}}$ can thus be viewed as a dynamical percolation threshold for fast growth reactions.  We indeed observe indications of such a cascading behavior for fast reactions at higher densities as discussed further below. In radical polymerization of dense, many chain systems with hindered termination this is probably related to the autoacceleration which can lead to a polymerization 'explosion'.~\cite{Achilias1992, Tran-Cong-Miyata2017}   \\

\section{Emerging properties from simulated chain growth}
\subsection{Size scaling for different reaction rates}

\begin{figure*}[!htbp]
\begin{center}
\includegraphics[width=17.8cm]{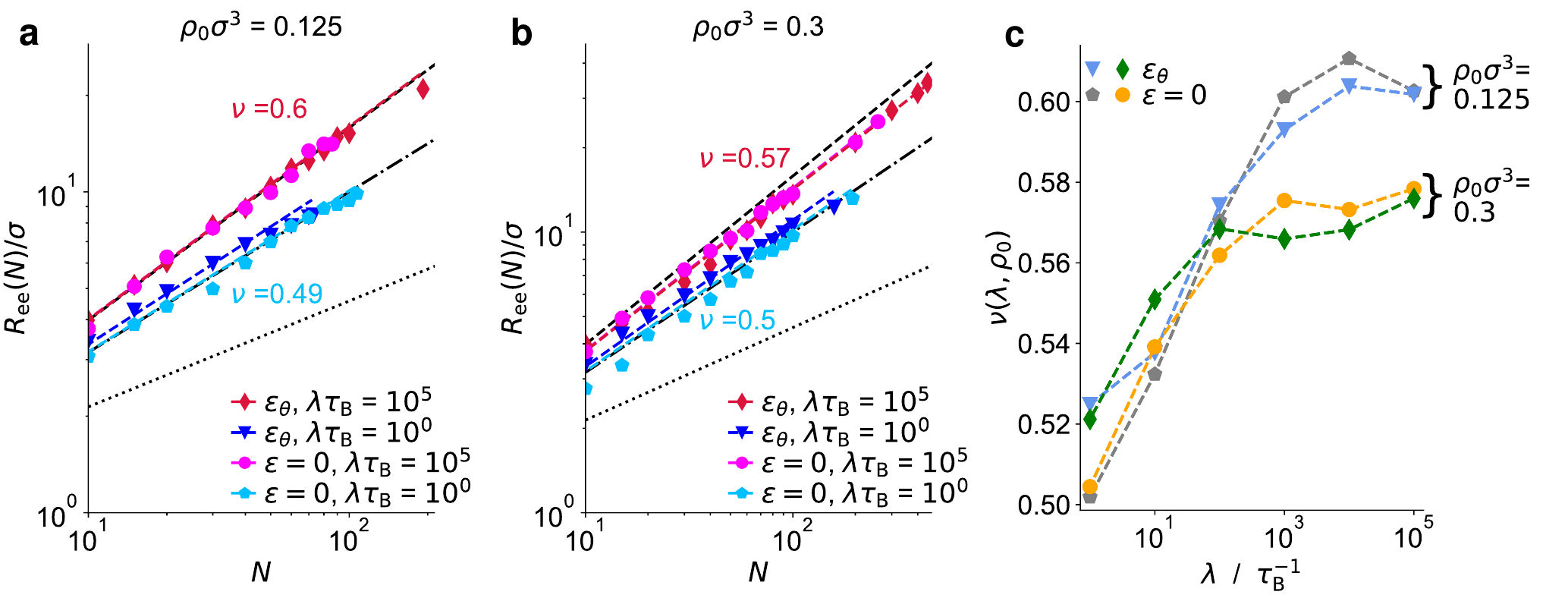}
\caption{\textbf{Scaling behavior of growing polymer chains.} (a) and (b) End-to-end distance for growing chains $R_{\mathrm{ee}}(N)$ in perfectly ideal ($\varepsilon=0$) and real-chain $\theta$ ($\varepsilon_\theta$) solvents for slow ($\lambda\tau_B=1$) and fast  ($\lambda\tau_B=10^5$) rates and for densities $\rho_{0} = 0.125$ and $0.3$~$\sigma^{-3}$, compared with the expected equilibrium behavior for good solvents $\nu = 3/5$ (black dashed), $\theta$-solvents $\nu = 1/2$ (black dash-dotted), and poor solvents $\nu = 1/3$ (black dotted). (c) Summary of scaling exponent $\nu (\lambda, \rho_{0})$ in dependence of the reaction propensity $\lambda$ for the two densities and the two solvent qualities as in (a) and (b).}
\label{Fig2:fig} 
\end{center}
\end{figure*}

The fast propagation of the chain leads to unexpected structural and conformational effects, in particular for how the polymer size changes with the degree of polymerization $N$ ('size scaling'). Fig.~\ref{Fig2:fig}a and b show end-to-end distances $R_{\mathrm{ee}}(N)$ of growing chains at densities $\rho_{0} = 0.125$ and $0.3$~$\sigma^{-3}$, respectively, in ideal and $\theta$-solvent conditions. The simulated data has been fitted with the typical power law $R_{\mathrm{ee}} = bN^{\nu}$ to obtain the scaling exponents $\nu$. In stark contrast to the equilibrium case, nonequilibrium growing chains in fast reaction conditions ($\lambda=10^5$) yield size scaling exponents representing a SAW scaling, that is, $\nu\simeq 3/5$. For lower propensities ($\lambda=10^0$), equilibrium scaling, $\nu\simeq 1/2$, is recovered. 

The dependence of $\nu$ with propensity $\lambda$ for the two densities  is summarized in Fig. \ref{Fig2:fig}c: clearly, low propensities $\lambda$ yield smaller exponents $\nu(\lambda, \rho_{0})$, and for both cases (fully non-interacting and $\varepsilon_{\theta}$) the random-walk value of $1/2$ is fully recovered. For larger propensites, $\lambda \gtrsim 10^2$, a transition to SAW scaling is observed for both ideal and $\theta$-conditions and both densities.  Note that the scaling exponent is lower for the higher density, $\rho_0\sigma^3=0.3$, although the total propagation rate is higher. This trend must be assigned to the existence of the dynamic percolation limit as predicted by mean field equation~(\ref{Rate:perc}), and which is approached here. Hence, the scaling exponents are depending non-monotonically on density. We will discuss this fact below again when we discuss the structural mechanisms and more generally a 'state' diagram of polymerization behavior versus $\lambda$ and $\rho_0$. 

\noindent

\subsection{Structural properties of chain and active center}

\begin{figure*}[!htbp]
\begin{center}
\includegraphics[width=17.8cm]{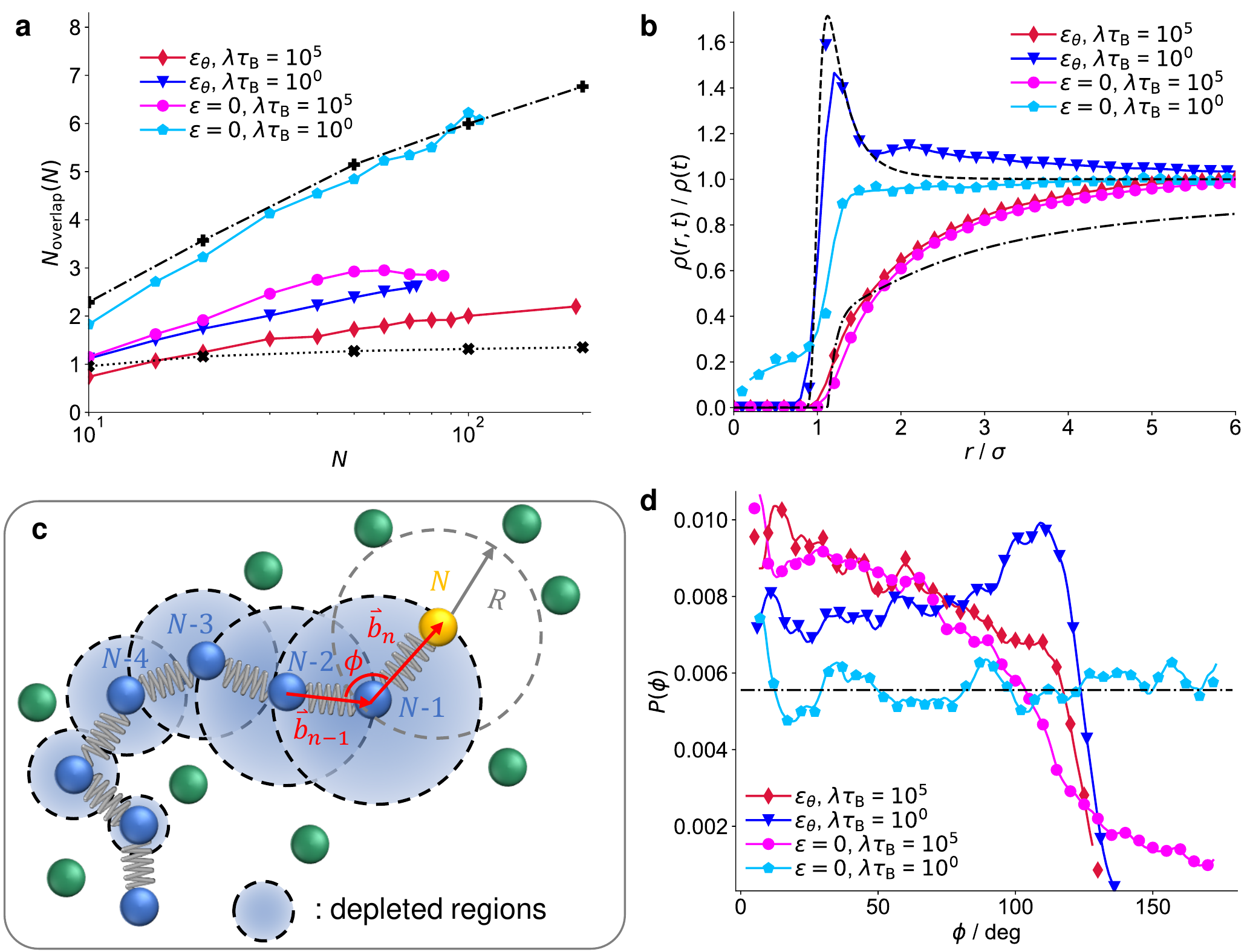}
\caption{\textbf{Structural nonequilibrium properties of growing polymer chains for ${\rho_{0} \sigma^{3} = 0.125}$}. (a) Number of non-bonded chain beads within a distance of $1.5$~$\sigma$ as measure for the overlap $N_{\mathrm{overlap}}(N)$. The nonequilibrium results for fast and slow growth are compared with the degree of overlap for equilibrium ideal chains (black, dashdotted line) and equilibrium chains with essentially repulsive, SAW-like, pairwise interactions of $\varepsilon = 0.1~k_{\rm B}T$ (black, dotted line). (b) Normalized radial densities profiles $\rho(r,t)/\rho(t)$ around the active center for different reaction conditions compared with the low-density equilibrium approximation of $g(r)$ for $\varepsilon_{\theta}$ (black, dashed line) and the theoretical profile in the steady-state diffusion-controlled ('Smoluchowski')  limit $\rho(r,t) = \rho(t)(1-R/r)$ (black, dashdotted line). (c) Schematic representation of directed growth of the (yellow) AC of a polymer chain away from the monomer-depleted zones (dashed-line circles) around the reacted blue polymer beads. The green spheres represent free monomers. (d) Probability distribution of bond angles $P(\phi)/\sin \phi$ between the new $\vec{b}_{n}$ and the previously formed bond vectors $\vec{b}_{n-1}$. The black, dashdotted line represents a homogeneous distribution among all possible angles.}
\label{Fig3:fig} 
\end{center}
\end{figure*}

The observed SAW-behavior indicates that for fast growth the chain is spatially self-avoiding, i.e., polymer beads do not overlap. In Fig.~\ref{Fig3:fig}a, we compare the average number of overlapping non-bonded chain beads $N_{\mathrm{overlap}}(N)$ from the growth simulations with their equilibrium counterparts for $\theta$-conditions and good solvent conditions (represented by $\varepsilon = 0.1~k_{\mathrm{B}}T$). This measure counts the number of overlaps within a threshold distance of 1.5~$\sigma$, and higher counts indicate more collapsed chain topologies. We indeed observe a lower degree of overlap and thus more extended chains for faster reaction conditions. This confirms the observations of Fig.~\ref{Fig2:fig} that fast growing chains are more swollen than they would be at equilibrium conditions. The nonequilibrium data for ideal and $\theta$-conditions is very well accommodated between the equilibrium ideal (upper) and equilibrium SAW (lower) bounds in Fig.~\ref{Fig3:fig}a. 

The self-avoidance can be explained by the nonequilibrium density profiles of free monomers around the ACs $\rho(r,t)/\rho(t)$ plotted in Fig.~\ref{Fig3:fig}b: They show a monomer-depleted zone at small distances $r$ even for a completely ideal system at high reaction rates. All monomers in that zone typically react before they approach closer, and thus an excluded volume - nonequilibrium correlation hole -  around the active center is created. For the fast reactions, the density profile around the classical Smoluchowski sink with $\rho(r) = 0$ for $r < R$ and $\rho(r)/\rho_{\infty} = 1 - R/r$ for all $r \geq R$ serves as a good orientation in the plot. For the slow reactions, no such correlation hole is observed, because the monomers have time to relax between reaction events. The profiles are then near equilibrium, as seen by comparing to the equilibrium radial distribution function in the low density limit being $\rho(r)/\rho_{\infty} = g(r) = r^{2} \exp[-\beta u_{\mathrm{LJ}}(r)]$ for the $\theta$-solvent. The presence of interactions leads to an excluded volume around each particle, which manifests at slower reaction in a lower degree of overlap in Fig.~\ref{Fig3:fig}a and $\rho(r,t) = 0$ for $r \lesssim \sigma$ in Fig.~\ref{Fig3:fig}b. Only in the ideal case for slow reactions, free monomers are found in very close proximity to the active center ($r < \sigma$), as expected for Doi-type of reactions~\cite{Dibak2019}.  

The depletion and local self-avoidance leads to a directed growth of the polymer chain, which rationalizes the ABP-like behavior and the 'persistence factor' $\alpha$ (and persistence time $\tau_{\rm rot}$) introduced in the jump diffusion theory above. The reason, illustrated in Fig.~\ref{Fig3:fig}c, is that a newly attached monomer, becoming the active center, has the density depletion hole in the region 'behind' it, i.e., at the location of the previous AC. This fact is evidenced by the probability distribution of growth angles $P(\phi)/\sin \phi$ (see Appendix \ref{SIsec3:sec}) in Fig.~\ref{Fig3:fig}d: A growth angle, defined by the angle between the previous and the newly formed AC bond, of $\phi = \pi=180^\circ$ degrees means that the growth is perfectly 'forward', whereas $\phi = 0$ means that the vectors point exactly in opposite directions. A flat distribution, i.e., all angles are equally likely and the chain grows randomly, is found in good agreement for the slow, ideal growth of a chain. However, for fast growth, angles larger than $\phi \approx 0.65 \pi = 116.5^{\circ}$ are much more unlikely than for slow growth. This exclusion confirms the concept of depleted zones in the wake of the moving active center. 

The observed effect of persistent self-avoiding motion is comparable to a clever worm which eats its way through an apple and which will always move in a direction where there is food, no voids. An APB analogy has been reported by~\citet{Merlitz2020}, where the orientation of the active particles without sensing the concentration gradient is driven by translational/directional memory, leading to pseudo-chemotaxis~\cite{Lapidus1980}.
However, in the polymerization case the food (free monomers) can move and diffuse. We actually observe that the depletion holes (cf. Fig.~\ref{Fig3:fig}b) for higher densities become much smaller (see Appendix \ref{SIsec5:sec}) since the percolating pathways and cascade-like reactions (very fast "sprints") are so fast that the monomers do not generate a stationary Smoluchowski density profile. This loss of depletion holes is the structural reason why scaling exponents decrease again for high densities (cf. Fig.~\ref{Fig2:fig}d). 

\noindent

\vspace{0.5cm}
\subsection{Rate and density dependence of scaling exponents}

The nonequilibrium scaling behavior is not limited to ideal or $\theta$-conditions. Fig.~\ref{Fig4:fig}a shows that the nonequilibrium effect on the scaling exponent $\nu(\varepsilon)$ is observed for all interactions $\varepsilon$, and thus fast growing polymer chains are more extended with respect to their equilibrium counterparts in all solvent conditions (see Appendix \ref{SIsec4:sec}). However, the effect is most substantial in ideal, $\theta$ and poor solvent conditions. As expected, for slow reactions the behavior transits back to the equilibrium situations. For discussing the conditions at which nonequilibrium effects occur more systematically, we screened the growth behavior for the ideal case for various densities ranging from $\rho_{0}\sigma^{3} = 0.01$ to $1.0$ and different $\lambda$ (see Appendix \ref{SIsec5:sec}). The resulting nonequilibrium 'state'-diagram is depicted in Fig.~\ref{Fig4:fig}b, representing random walk, intermediate, and SAW-scaling behavior: slow reactions (low $\rho_{0}$ and $\lambda$) yield the equilibrium result of $\nu \approx 0.5$ for $\theta$-solvents; increasing $\rho_{0}$ and $\lambda$ causes faster reactions which results in rising values of the scaling exponent closer to a good solvent behavior. For very large densities, roughly beyond the critical Smoluchowski percolation density $\rho \gtrsim \rho_{\mathrm{crit}} \approx 0.43$~$\sigma^{-3}$ (see Eq.~(\ref{Rate:perc})),  the exponents decrease again, as argued above. 

\section{Criteria for time-controlled solvent quality}

To understand the topology of the $\lambda-\rho_{0}$ state diagram theoretically, let us more deeply inspect the time scales in these systems: we have the 'process' time scale of reaction $\tau_{\mathrm{react}} = (\rho k(\lambda, \rho)^{-1}$, the intrinsic times scales of free monomer diffusion, $\tau_B = \sigma^2/D_0$ (equivalent to $b^2/D_0$ in our model), and finally the time scales of polymer segment relaxation $\tau_{p} = b^{2}/(6\pi^{2}D_0) s^2$ for segment size $N/p = s$. As argued before, depletion correlations and persistent directed growth should be visible if the reaction time scale is faster than other relevant relaxation time scales. For Smoluchowski-type of depletion holes to be present (recall Fig.~\ref{Fig3:fig}b), the reaction rate should be not slower than the diffusion time scale, i.e., $\tau_{\mathrm{react}} \lesssim \tau_B$. Comparing $\tau_B$ and Rouse segment relaxation, $\tau_p$, for this condition, we find that segments of the lengths $s \gtrsim 8$ cannot relax within 2 reaction events. Hence, we see that if condition  $\tau_{\mathrm{react}} \lesssim \tau_B$ is fulfilled, according to Rouse, also $\tau_{\mathrm{react}} \lesssim \tau_p$ for a sizeable segment length of a minimum of $n=8$ is fulfilled. Hence, very universally a growing segment of a few monomers remains persistent during the (fast) reaction time scale, as indeed observed in our structural analysis.

\begin{figure*}[!htbp]
\begin{center}
\includegraphics[width=17.8cm]{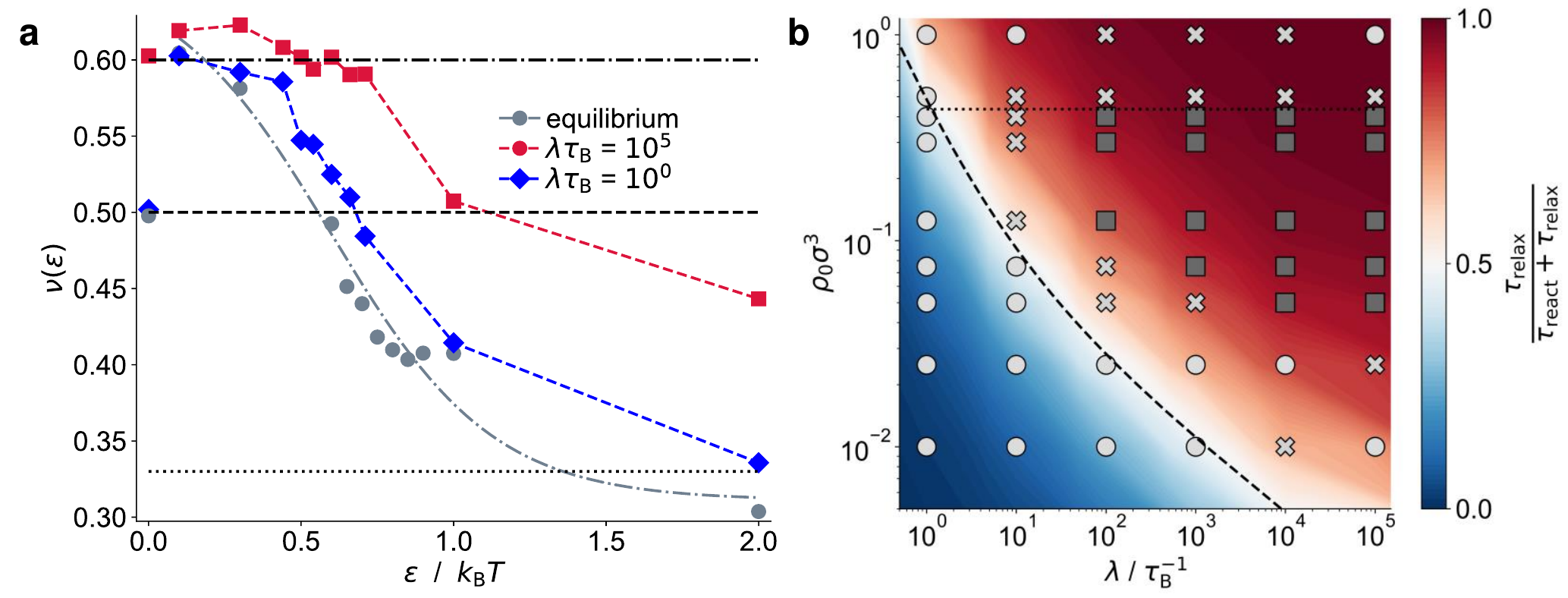}
\caption{\textbf{Summary of conformational scaling and state diagram.} (a) Scaling exponent $\nu$ as a function of the solvent quality (expressed by interaction paramer $\varepsilon$) for fast growing (red square), slowly growing (blue diamond), and  non-growing polymer chains in equilibrium (gray circle), all at $\rho_{0}\sigma^{3} = 0.125$. The black dash-dotted, dashed, and dotted lines serve as orientation for reference exponents $\nu = 3/5$, $1/2$, and $1/3$, respectively. (b) State diagram for growing ideal polymer chains ($\varepsilon =0$). The circles, crosses, and squares represent exponents $\nu < 0.525$ ($\theta$-solvent), $0.525 \leq \nu < 0.55$ (intermediate), and $\nu \geq 0.55$ (good solvent), respectively. The smooth contours in the back represent the ratio between reaction time scale $\tau_{\mathrm{react}}$ and the segmental relaxation $\tau_{\mathrm{relax}}$ time of a segment with $N/p^{*} =s^{*} = 5$. Red regions illustrate chain relaxation times $\tau_{\mathrm{relax}} \gg \tau_{\mathrm{react}}$, while blue regions depict $\tau_{\mathrm{relax}} \ll \tau_{\mathrm{react}}$. The bright-white thin region in between, together with the black dashed line corresponds to the boundary $\tau_{\mathrm{relax}} = \tau_{\mathrm{react}}$. The horizontal black dotted line represents the critical Smoluchowski percolation density $\rho_{\mathrm{crit}} = 0.43$ $\sigma^{-3}$, as predicted from the self-consistent Smoluchowski approach, Eq.~(\ref{Rate:perc}).}
\label{Fig4:fig} 
\end{center}
\end{figure*}

From these arguments we can estimate a theoretical threshold where nonequilibrium effects should start to play a role:  The coloring of the background contours in Fig.~\ref{Fig4:fig}b represents a comparison between the relaxation time $\tau_{\mathrm{relax}}=\tau_{p^{*}}$ of a Rouse segment with $N/p^{*} = s^{*} = 5$ and the reaction scale $\tau_{\mathrm{react}}$, using the ratio $\tau_{\mathrm{relax}}/(\tau_{\mathrm{relax}}+ \tau_{\mathrm{react}})$. The fit of the segment size of 5 beads provides the best agreement with the observations from simulations. It is indeed close to the analytical estimate made above.  The larger this ratio $\tau_{\mathrm{relax}}/(\tau_{\mathrm{relax}}+ \tau_{\mathrm{react}})$ is, e.g., approaching unity, the faster the reaction is compared to the segment relaxation and we observe nonequilibrium behavior. If the ratio is $\ll 1$, the segments relax much quicker than new elements are added and the growing chain exhibits equilibrium random walk behavior. The black dashed line in Fig.~\ref{Fig4:fig}b, which represents the combinations of $\lambda$ and $\rho_{0}$, where $\tau_{\mathrm{relax}} = \tau_{\mathrm{react}}$, that is, $\tau_{\mathrm{relax}}/(\tau_{\mathrm{relax}}+ \tau_{\mathrm{react}})=0.5$, represents in good agreement with the simulation data the topology of the boundary between equilibrium and nonequilibrium behavior. 

Finally, it is also important to briefly note on another time scale in the problem: the observation (simulation) time scale $\tau_{\mathrm{sim}}$, as often considered as the 'Deborah number' in polymer rheology~\cite{Reiner1964}. If this observation window is long and thus the polymer can relax entirely, also equilibrium scaling of the total chain size should be approached. The simulation time scale can be related with the mean number of added monomers $N$ as $\tau_{\mathrm{sim}} = N\tau_{\mathrm{react}}$. Comparing this polymerization time to polymer segment relaxation of segment length $s$, $N\tau_{\mathrm{react}} = b^{2}/(6\pi^{2}D_{\mathrm{0}})s^{2}$ with $\tau_{\mathrm{react}} \lesssim \tau_{\mathrm{B}} = b^{2}/D_{\mathrm{0}}$ as the diffusive time scale, we find that segments of size $s \lesssim 8 \sqrt{N}$ have maximally relaxed within the time simulated. Hence, for fast reactions, the full chain will never relax in the observation window but only segments with a {\it relative} amount of the chain decreasing as $\propto 1/\sqrt{N}$.  This nonequilibrium relaxation of polymer chains has analogies to nonequilibrium relaxation of chains after releasing them from an applied force $f$~\cite{DeGennes1979, Sheng1997}, only that in the case of polymerization this relaxation will propagate spatially from the beginning (initially polymerized segments) to the growing end, i.e., exhibits strong spatiotemporal heterogeneity.

\noindent

\section{Discussion}

Our simulations provide unprecedented molecular-level insight into the nonequilibrium structuring of polymers during chain-growth polymerization.  In particular, a directed single-chain polymerization appears with increased size scaling exponents at faster reaction conditions. This effect originates from the coupling and competition of timescales and the resulting spatio-temporal correlations, even in the ideal, fully noninteracting case. Our results complement known concepts of polymerization kinetics for termination rates of many-chain systems, where the chain growth is characterized by 'reaction diffusion' jumps and the Trommsdorff-Norrish-effect~\cite{Schulz1956, DeGennes1982a, Russell1988, Achilias1992}, by novel structural insights. Our insights apply to polymerization on all scales, for example, self-assembly of patchy, colloidal polymers~\cite{Sciortino2007, Lu2008, Groeschel2013}.

The growth of rather stretched chain morphologies at fast, nonequilibrium conditions compared to collapsed structures in equilibrium at $\theta$- or poor solvent conditions opens pathways towards the design and the creation of novel adaptive materials~\cite{Liu2018, Walther2019} for industrial applications, potentially using fast 3D and 4D (photo)polymerization techniques~\cite{Muelhaupt2017, Bagheri2019, Telitel2020}. During nonequilibrium processing, tensely stretched chain segments may contain anisotropic stress and stored energy over their equilibrium counterparts, and controlled storage and delayed energy release back to equilibrium will constitute fascinating aspects for future studies~\cite{Chandran2019, Reiter2020}. Experimental evidence of uncontrolled release of stretching energy can be found in material shrinkage during and after autoacceleration~\cite{Tran-Cong-Miyata2017}. %

Interestingly, the active centers in fast chain polymerizations behave analogously to APBs in search of food, but in contrast to APBs this originates from the synergistic physics of the growing chain and diffusive monomer motion rather than the properties of the single particles. Due to the analogy, it will be interesting to investigate if polymerization in dense, many chain systems shows collective behavior similarly rich as for ABPs, such as 'swarming', microclustering, ratchet-effects, activity-induced phase transitions, or novel interfacial phenomena~\cite{Bechinger2016}. Such spatiotemporal pattern formation is very typical in reaction-diffusion processes and, if controllable, can lead to novel hierarchical material structures on various length and time scales. 

\section{Methods}
\subsection{Brownian Dynamics}
We performed all simulations in the LAMMPS software package~\cite{Plimpton1995} using the formalism for overdamped Langevin dynamics ('Brownian Dynamics' - BD). The inertia-free Langevin equation of motion for a particle $i$ writes
\begin{equation}
\xi_{i}\mathbf{\dot{r}}_{i} = - \nabla U(\mathbf{r}_{i}) + \mathbf{R}(t) \;, 
\label{BD:eq1}
\end{equation}
where $\mathbf{\dot{r}}_{i}$ and $\mathbf{r}_{i}$ denote velocity and position of the $i$-th particle, and the drag coefficient $\xi_{i}$ and the diffusion coefficient $D_{i}$ are related through the Einstein relation $D_{i} = k_{\mathrm{B}}T/\xi_{i}$, and $\mathbf{R}(t)$ is the random force vector. All diffusion coefficients $D_{i}$ are set to $D_{0}\tau_{B}/\sigma^{2} = 1.0$ with $\sigma = 1.0$ as the van der Waals radius of the particles as well as our unit for length. The components of the random force vector fulfill the properties $\langle R_{\alpha}(t) \rangle = 0$ and $\langle R_{\alpha}(t) R_{\beta}(t^{'})\rangle = 2\xi_{i}^{2}D_{i}\delta_{\alpha \beta} \delta(t - t^{'})$ with $\alpha$ and $\beta$ denoting the spatial dimensions, and $\delta$ being the Dirac $\delta$-function. The pair-wise non-bonded interactions are described through a 12-6-Lennard-Jones potential, which writes
\begin{equation}
u_{\mathrm{ij}}^{\mathrm{LJ}}(r_{ij}) = 4 \varepsilon\left[\left(\frac{\sigma}{r_{ij}} \right)^{12} - \left( \frac{\sigma}{r_{ij}}\right)^{6} \right] \; ,
\end{equation}
where $r_{ij}$ is the distance between the particles $i$ and $j$, and $\varepsilon$ as the depth of the potential well determines the strength and nature of the interactions as well defines our solvent quality (see Appendix \ref{SIsec1:sec}). Between bonded neighbors $i$ and $j$ in the polymer chain, we only apply a harmonic bond potential with $u_{ij}^{\mathrm{bond}}(b_{ij})= K_{\mathrm{bond}}(b_{ij}-b_{0})^{2}$, where $b_{ij}$ is magnitude of the bond vector, $K_{\mathrm{bond}} = 20 \varepsilon \sigma^{-2}$ is the spring constant, and $b_0=\sigma$ is the equilibrium bond length. For a free monomer $i$, the corresponding force acting on it writes
\begin{equation}
\mathbf{F}_{i} = -\nabla U(\mathbf{r}_{i}) = -\sum_{i \neq j}^{N_{\mathrm{tot}}} \nabla u_{ij}^{\mathrm{LJ}}(r_{ij}) \; , 
\end{equation}
where $N_{\mathrm{tot}}$ is the total number of particles in the system and for $n$-th (non-terminal) chain bead
\begin{equation}
\begin{split}
\mathbf{F}_{n} &= - \nabla u^{\mathrm{bond}}_{n-1,n}(b_{n-1,n}) + \nabla u^{\mathrm{bond}}_{n,n+1}(b_{n,n+1}) \\ &-  \sum_{j \not\in \{n - 1, n, n + 1\}}^{N_{\mathrm{tot}}} \nabla u_{nj}^{\mathrm{LJ}}(r_{nj}) \; ,
\end{split}
\end{equation}
where for the terminal beads $n = 1$ and $N$, the first or second bonded contribution are omitted, respectively. The positions of all $N$ particles are updated using the Euler-Maruyama propagation scheme \cite{Ermak1978}, which writes
\begin{equation}
\begin{split}
\mathbf{r}_{i}(t + \Delta t) = \mathbf{r}_{i}(t) + \frac{\Delta t}{\xi_{i}} \mathbf{F}_{i} + \sqrt{2D_{0} \Delta t} \bm{\zeta}_{i} \; , 
\end{split}
\end{equation}
where $\Delta t$ is the integration time-step, which is $10^{-5} \tau_{B}$ for all systems investigated, and $\bm{\zeta}_{i}$ is a vector consisting of random values following a standard normal distribution. \\

\subsection{Bond formation}
Every $\lambda \Delta t$ integration time steps, the algorithm checks if a bond formation is possible with a probability $p_{\mathrm{react}} = 1.0$. In the simulations, reaction checks are performed between the nearest free monomer and the active center using the cut-off protocol method~\cite{Akkermans1998, Farah2012, DeBuyl2015} following the Doi scheme \cite{Doi1975a, Erban2009, Dibak2019} using the LAMMPS implementation~\cite{DeBuyl2015} for creating new bonds within a spherical reactive volume with a reactive radius $R = \sqrt[6]{2}\sigma \ \approx 1.122 \sigma$ around the AC (see Fig.~\ref{Fig1:fig}a). Once a new bond is formed, the properties of the AC are transferred to the newly added monomer and the previous active center is deactivated. \\

\subsection{Simulation details}
The cubic simulation box with periodic boundary conditions initially contains $1000$ non-reacted (free) monomers and a single AC particle as seed for the growing polymer chain. Different number densities $\rho_{0} = N_{\mathrm{m},0}/V$ are studied by varying the simulation cell's volume $V$ and keeping the total initial amount of particles $N_{\mathrm{m},0}$ constant. We conducted simulations for observation time windows up to $2 \cdot 10^{3}\tau_{\mathrm{B}}$ for densities $\rho_{0}\sigma^{3} = \{0.01, \, 0.025, \, 0.05, \, 0.075, \, 0.125, \, 0.3, \, 0.4, \, 0.5, \, 1.0\}$ for $\varepsilon = 0$. Systems with pairwise interactions of strength $\varepsilon/(k_{\mathrm{B}}T) = \{ 0.1, \, 0.3, \, 0.44, \, 0.5, \, 0.54, \, 0.6, \, 0.66, \, \, 0.71, \, 1.0, \, 2.0 \}$ have been simulated for the density $\rho_{0} = 0.125$. For $\rho_{0}\sigma^{3} = 0.3$, systems with $\varepsilon/(k_{\mathrm{B}}T) = {0.44, \, 0.54, \, 0.71, \, 0.8, \, 0.9}$ were studied. All simulations with reaction propensities ranging from slow $\lambda = 10^{0}$ to fast $\lambda = 10^{5}$~$\tau_{\mathrm{B}}^{-1}$ have been conducted with a self-written wrapper around the LAMMPS software package \cite{Plimpton1995}. The amount of different simulated trajectories for a combination of $\varepsilon$, $\lambda$, and $\rho_{0}$ ranges from $10^{1}$ to $10^{3}$ independent runs. For comparison with the equilibrium state and for determining a suitable value for $\varepsilon_{\theta}$, simulations of non-growing polymer chains of lengths $N = \{10, \, 20, \, 50, \, 100, \, 150, \, 200\}$ have been performed for a broad range of interaction parameters with monomers present at densities $\rho_{0}\sigma^{3} = 0.125$ and $0.3$ (see Appendix \ref{SIsec4:sec} and \ref{SIsec5:sec}).

\begin{acknowledgements}
The authors acknowledge support by the state of Baden-Württemberg through bwHPC and the German Research Foundation (DFG) through grant no INST 39/963-1 FUGG (bwForCluster NEMO). The authors thank Benjamin Rotenberg, Stefano Angioletti-Uberti, Sebastian Milster, Matthias Ballauff, and Günter Reiter for inspiring discussions and useful comments.
\end{acknowledgements}
\appendix

\section{Determination of $\varepsilon_{\theta}$ from chains in equilibrium}
\label{SIsec1:sec}
This section contains a justification for the choice of the interaction parameter for $\theta$-conditions $\varepsilon_{\mathrm{\theta}}/(k_{\mathrm{B}}T) = 0.6$ and $0.8$ for densities $\rho_{0}\sigma^{3} = 0.125$ and $0.3$, respectively, used throughout our work for the Lennard-Jones pairwise interactions in a Browian dynamics (BD) framework. To represent $\theta$-conditions in our simulations, we needed to find a value for $\varepsilon$, where the excluded volume $v$ of the chain beads and the surrounding solvent vanishes with $v = 0$ \cite{Rubinstein2003}. It is not sufficient to chose $\varepsilon$ such that the second virial coefficient $B_{2} = 0$ describing the nature of the pairwise interactions, since a correction considering three-body effects needs to be considered for polymer chains \cite{DeGennes1979, Heyda2013}. There are many definitions of the $\theta$-point \cite{Zhang2020}, but literature reports generally values between $\varepsilon/k_{\mathrm{B}}T = 0.25$ and $0.5$ for MD simulations \cite{Ciesla2007}, for the Langevin framework \cite{Heyda2013}, and for MC simulations \cite{Rubio1995, Yong1996, Graessley1999} in the absence of monomers or co-solvents. It has been reported that chain-length dependent size effects appear, which require an increased value of $\varepsilon_{\theta}$ for longer chains $N$ \cite{Milchev1993, Yong1996, Graessley1999}. \citet{Zhang2020} presents four different special temperatures in the $\theta$-regime, which all differ for non-infinite chain lengths $N$. \\

However, except for the work by \citet{Heyda2013}, the influence of explicit solvent molecules in form of co-solutes on the determination of the interactions describing the $\theta$-point has not been reported to our knowledge. Therefore, we decided to determine $\varepsilon_{\theta}$ for two densities $\rho_{0}\sigma^{3} = 0.125$ and $0.3$ from equilibrium simulation without chain growth steps including free monomers as co-solutes. The total amount of particles $N_{\mathrm{tot}} = N + N_{\mathrm{mono}}$ with $N_{\mathrm{mono}}$ as the amount of monomers remained constant, while the volume was changed to obtain the two different densities. All pairwise interactions between all non-bonded pair of all species were described by a single radius $\sigma$ and a single $\varepsilon$. \\

\begin{figure}[!htbp]
\begin{center}
\includegraphics[width=8.6cm]{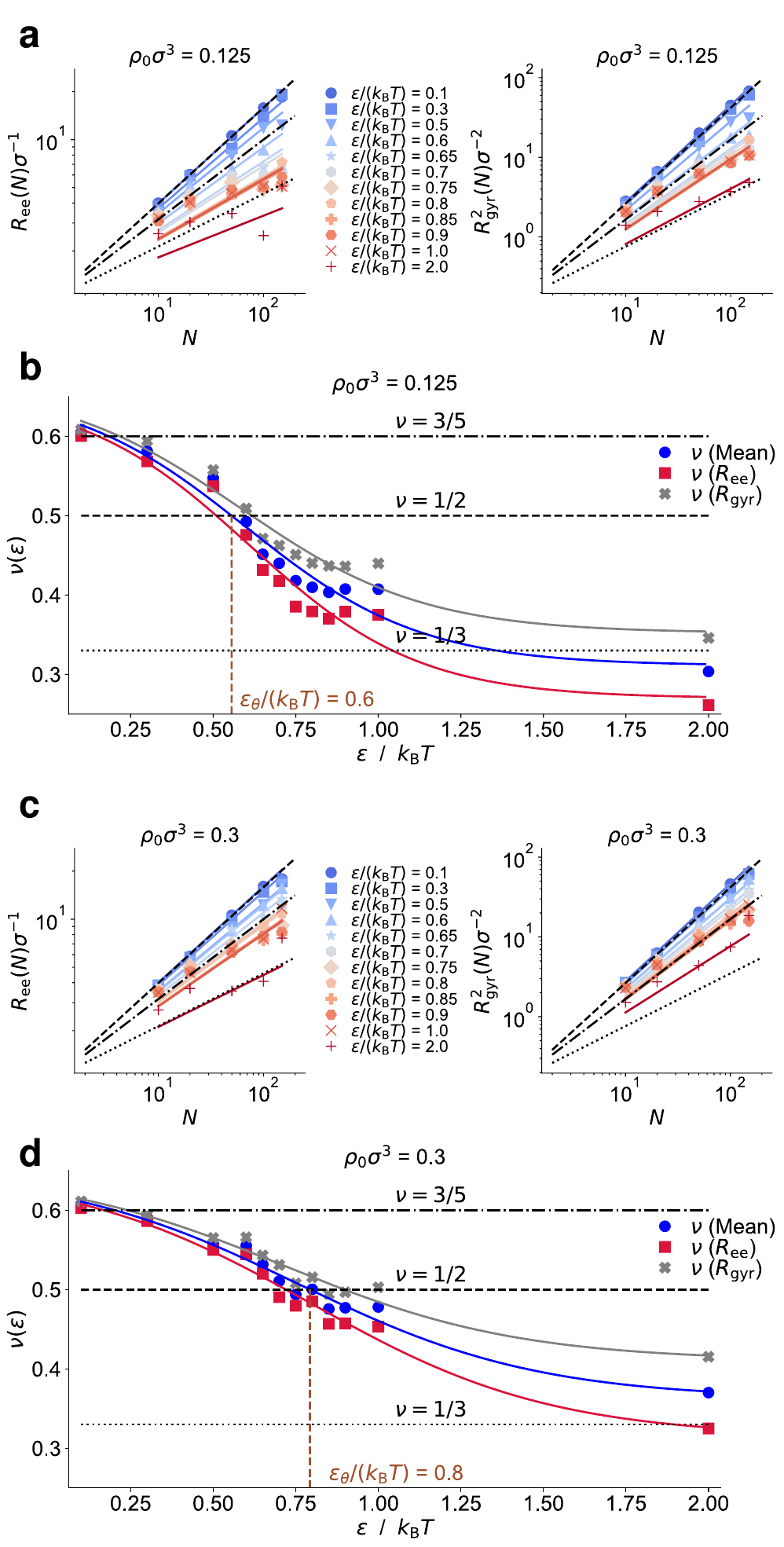}
\caption{\textbf{Determiniation of the $\theta$-point for different $\varepsilon$ for different densities $\rho_{0}$.} (a) and (b) show the mean simulated end-to-end distances $R_{\mathrm{ee}}(N) \propto N^{\nu}$ and the squared radius gyration $R_{\mathrm{gyr}}^{2}(N) \propto N^{2\nu}$ for $\rho_{0}\sigma^{3} = 0.125$ and $0.3$, respectively. Symbols represent simulation results for a given chain length $N$ and interaction parameter $\varepsilon$. Lines represent the corresponding fits following Equation \ref{ReeRgyr:SIeq}, respectively. Dotted, dashed, and dash-dotted line represent theoretical scaling behavior for bad ($\nu = 1/3$), $\theta$ ($\nu = 1/2$), and good solvent conditions ($\nu = 3/5$), respectively. (b) and (d) show the $\varepsilon$-dependency of the scaling exponent. Circles, squares, and crosses represent fitted scaling exponents $\nu(\varepsilon)$ (Equation \ref{ReeRgyr:SIeq}) from simulated data  for $R_{\mathrm{ee}}$ (red), $R^{2}_{\mathrm{gyr}}$ (gray), and the mean of the two (blue), respectively. The lines represent the corresponding sigmoidal fits using Equation \ref{SigFit:SIeq}. Values for $\varepsilon_{\mathrm{\theta}}$ are represented by the brown dashed lines. Black dotted, dashed, and dash-dotted line represent theoretical scaling behavior for bad ($\nu = 1/3$), $\theta$ ($\nu = 1/2$), and good solvent conditions ($\nu = 3/5$), respectively.}
\label{Fig1:SIfig} 
\end{center}
\end{figure}

$\varepsilon_{\theta}$ is determined from equilibrium Brownian dynamics (BD) simulations of non-growing polymer chains of lengths $N = $ $10$, $20$, $50$, $100$, $150$ and a corresponding amount of interacting monomer particles $N_{\mathrm{mono}} = $ $990$, $980$, $950$, $900$, $850$ in the presence of pair-wise non-bonded interactions with $\varepsilon/(k_{\mathrm{B}}T) = $ $0.1$, $0.3$, $0.5$, $0.6$, $0.65$, $0.7$, $0.75$, $0.8$, $0.85$, $0.9$, $1.0$, $2.0$. The increasing value of $\varepsilon$ corresponds to a decreasing temperature with $\varepsilon \propto T^{-1}$, and larger values of $\varepsilon$ lead to more attractive interactions and a globular chain. The smaller $\varepsilon$ is, the more extended the chains are. For each combination of $N$, $\rho_{0}$ and $\varepsilon$ up to 30 different trajectories have been simulated and analyzed. Data was collected after equilibrating for at least $5\tau_{\mathrm{Rouse}}$ with the Rouse time $\tau_{\mathrm{Rouse}} \propto N^{2}$ to ensure a sufficient relaxation of the chain structures. The obtained results for the end-to-end distance (ee) $R_{\mathrm{ee}}$ and the radius of gyration (gyr) $R_{\mathrm{gyr}}$ are depicted in Figs. (\ref{Fig1:SIfig}a) and (\ref{Fig1:SIfig}c) for densities $\rho_{0}\sigma^{3} = 0.125$ and $0.3$, respectively. For determining the universal scaling exponent $\nu$ of the polymer size with $R \propto N^{\nu}$ \cite{Flory1953, DeGennes1979, Doi1986, Rubinstein2003}, both $R_{\mathrm{ee}}$ and $R_{\mathrm{gyr}}$, have been fitted (lines in Fig. (\ref{Fig1:SIfig}a) and (\ref{Fig1:SIfig}c) based on 
\begin{equation}
R_{\mathrm{ee}}^{2} = b^{2} N^{2\nu} = 6 R_{\mathrm{gyr}}^{2} \; ,
\label{ReeRgyr:SIeq}
\end{equation}
where $b$ is the equilibrium bond length. Figs. (\ref{Fig1:SIfig}b) and (\ref{Fig1:SIfig}d) show the scaling behavior of polymer chains for various $\varepsilon$ for $\rho_{0}\sigma^{3} = 0.125$ and $0.3$. Interactions of repulsive nature with $\varepsilon/(k_{\mathrm{B}}T) = 0.1 - 0.3$ yield good solvent behavior with $\nu \approx 3/5$ independent of the surrounding monomer concentration. Bad solvent behavior is observed for more attractive values of $\varepsilon/(k_{\mathrm{B}}T) \geq 0.7$ for $\rho_{0}\sigma^{3} = 0.125$. At higher densities ($\rho_{0}\sigma^{3} = 0.3$ - Fig.~(\ref{Fig1:SIfig}d)), free monomers stabilize the more extended chain configurations and therefore higher scaling exponents are found for the same values of $\varepsilon$ compared to the lower densities. Only for the lower density the characteristic bad solvent scaling behavior for higher $\varepsilon$ is observed and the chain collapses to a globular state. \\

All fitted $\nu_{\mathrm{ee}}$ and $\nu_{\mathrm{gyr}}$ from Figs. (\ref{Fig1:SIfig}a) and (\ref{Fig1:SIfig}c) are shown together with the mean $\nu_{\mathrm{mean}} = 1/2(\nu_{\mathrm{ee}} + \nu_{\mathrm{gyr}})$ in Figs. (\ref{Fig1:SIfig}b) and (\ref{Fig1:SIfig}d), respectively. The scaling exponents $\nu(\varepsilon)$ are fitted with a sigmoidal fit (lines in Figs. \ref{Fig1:SIfig}b and c) of the form
\begin{equation}
\nu_{\mathrm{fit}}(\varepsilon) = A + \frac{B}{C + \exp(-D\varepsilon)} \, ,
\label{SigFit:SIeq}
\end{equation} 
where $A$, $B$, $C$, and $D$ represent fitting parameter to interpolate between the simulated data points in good agreement. Solving Equation \ref{SigFit:SIeq} for $\nu_{\mathrm{fit}} = 0.5$ then yields $\varepsilon_{\theta}/(k_{\mathrm{B}}T) \approx 0.6$ and $0.8$ for $\rho_{0}\sigma^{3} = 0.125$ and $0.3$, respectively. For $\nu(R_{\mathrm{ee}})$ at high and low densities, $\varepsilon_{\mathrm{\theta}}/(k_{\mathrm{B}}T) = 0.5$ and $0.7$ are obtained, respectively, and for $\nu(R_{\mathrm{gyr}})$ $0.6$ and $0.9 k_{\mathrm{B}}T$ are found with respect to the corresponding fits.  
\noindent

Fig. (\ref{Fig3:fig}a) addresses the degree of overlap with respect to the chain length, and we compared the number of non-bonded neighbors around the chain beads for selected non-equilibrium simulations in comparison with ideal chains and $\varepsilon/(k_{\mathrm{B}}T) = 0.1$. Simulations with this value always yield scaling exponents $\nu \approx 3/5$ representing the good solvent limit.

\section{Definition of the probability distribution of angles $P(\phi)$}
\label{SIsec3:sec}
\begin{figure}[!htbp]
\begin{center}
\includegraphics[width=8.6cm]{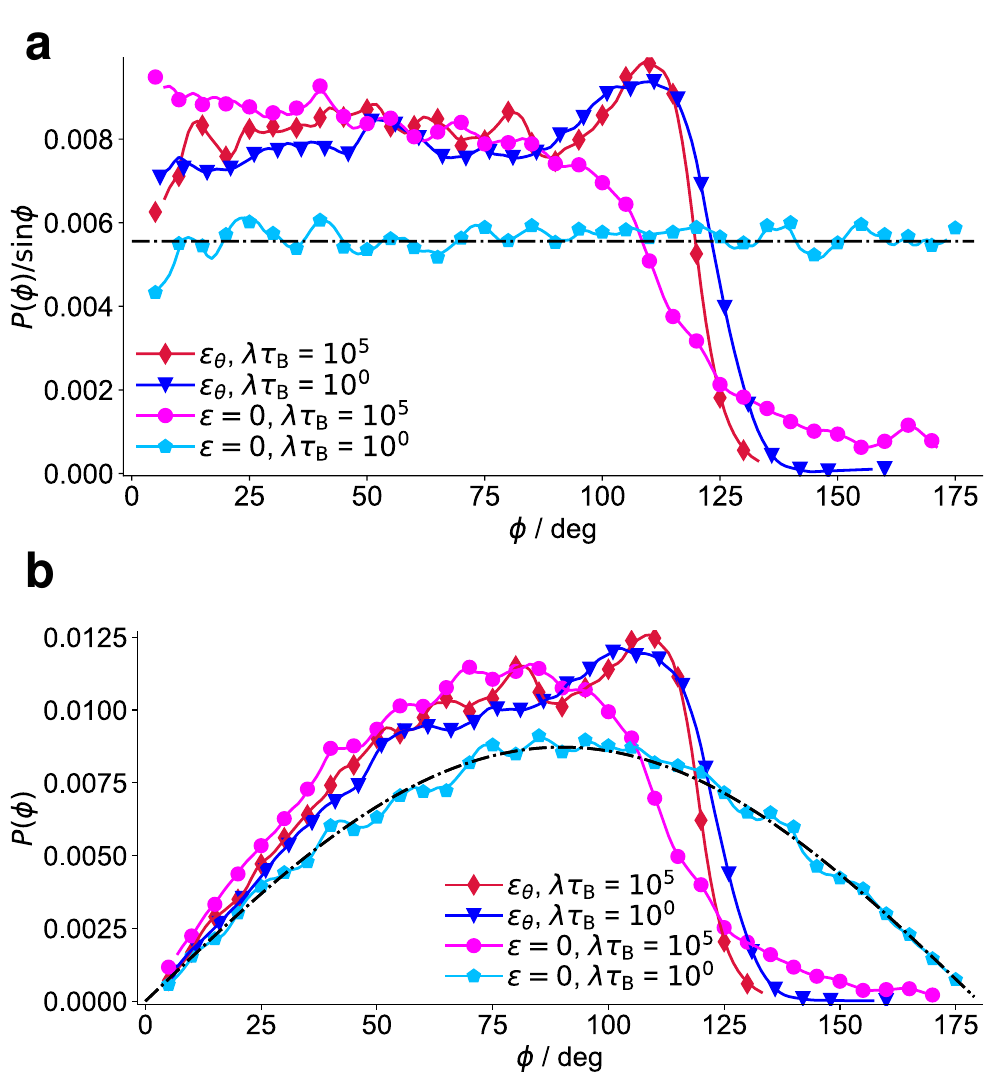}
\caption{\textbf{Angle distributions at the growing end of the chain for $\rho_{0}\sigma
^{3} = 0.3$.} (a) Normalized probability distribution $P(\phi)/\sin \phi$ and (b) effective probability distribution $P(\phi)$ for angles $\phi$ between the latest two added bonds of a growing chain. The dashdotted, black lines represent the homogeneous distribution among all possible angles.}
\label{Fig6:SIfig} 
\end{center}
\end{figure}

There exist multiple possibilities to represent the probability distribution of newly formed bond angles spanned up by the last two bond vectors $\vec{b}_{n}$ and $\vec{b}_{n-1}$. Figures~(\ref{Fig6:SIfig}a) and (\ref{Fig6:SIfig}b) show the normalized $P(\phi)/\sin \phi$ and the original probability distribution $P(\phi)$ of these angles. The black dashdotted line in Fig.~(\ref{Fig6:SIfig}a) is given by $\frac{1}{\sum_{\phi = 0}^{\pi} \sin \phi}$ and the homogeneous distribution as represented by the black, dashdotted line in Fig.~(\ref{Fig6:SIfig}b) is $\sin \phi/(\sum_{\phi =0}^{\pi} \sin \phi)$.  

\section{Non-equilibrium growth behavior for equilibrium good and poor solvent $\varepsilon$}
\label{SIsec4:sec}
\begin{figure}[!htbp]
\begin{center}
\includegraphics[width=8.6cm]{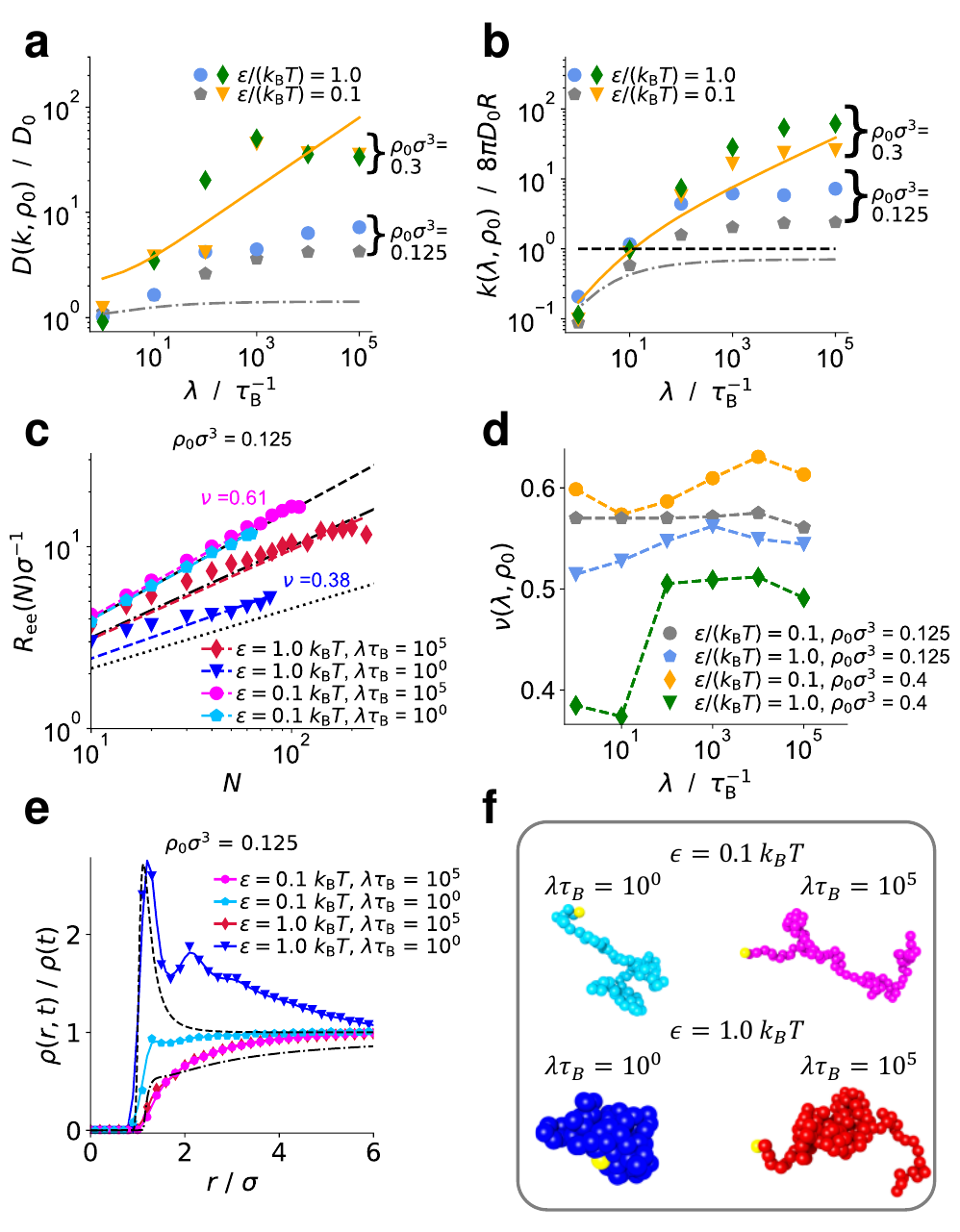}
\caption{\textbf{Growth at good ($\varepsilon = 0.1~k_{\mathrm{B}}T$) and poor solvent conditions ($\varepsilon = 0.1~k_{\mathrm{B}}T$) for densities $\rho_{0}\sigma
^{3} = 0.125$ and $0.4$.} (a) Simulated diffusion coefficients $D$ (symbols) for different reaction rates $k$ and initial densities $\rho_{0}$. The orange line and the dashdotted gray line represent the fastest and slowest theoretical limit for $\rho_{0}\sigma^{3} = 0.4$ and $0.125$, respectively. (b) Simulated reaction rate constants $k$ (symbols) for different reaction propensities $\lambda$ and initial densities $\rho_{0}$. The orange line and the dashdotted gray line represents the fastest and slowest theoretical limit for $\rho_{0}\sigma^{3} = 0.4$ and $0.125$, respectively, and the dashed black line the fastest Smoluchowksi reaction rate $k_{\mathrm{S}} = 8 \pi D_{0}R$. (c) End-to-end distance at a density $\rho_{0}\sigma^{3} = 0.125$ for good and poor solvent conditions during fast and slow reactions. Symbols show selected simulated data points, colored dashed lines represent corresponding fits with $R_{\mathrm{ee}}(N) \propto N^{\nu}$. Black dashdotted, dashed, and dotted lines represent theoretical curves for $\nu = 3/5$ (good solvent), $\nu = 1/2$ ($\theta$-solvent), and $\nu = 1/3$ (poor solvent). (d) Size scaling exponents $\nu$ for different propensities $\lambda$. (e) Normalized densities profiles $\rho(r,t)/\rho(t)$ around the active site for different reaction conditions compared with the theoretical profile in the diffusion-controlled ('Smoluchowski') limit $\rho(r,t) = \rho(t)(1-R/r)$ (black, dashdotted line)~\cite{Smoluchowski1918, Dibak2019}. (f) snapshots of chains at different solvent conditions for a density $\rho_{0}\sigma^{3} = 0.125$ at fast ($\lambda\tau_{\mathrm{B}} = 10^{5}$) and slow ($\lambda\tau_{\mathrm{B}} = 10^{0}$) reaction conditions with yellow-colored active sites.}
\label{Fig4:SIfig} 
\end{center}
\end{figure}

Fig. (\ref{Fig4:fig}a) of the article shows that enhanced scaling exponents $\nu$ are also found for other values of the interaction strength $\varepsilon$. Especially for $\varepsilon >\varepsilon_{\theta}$, Fig. (\ref{Fig4:fig}a) demonstrates a drastic shift from poor solvent size scaling $\nu \approx 1/3$ and collapsed coils which would be expected for this $\varepsilon$ in Equilibrium (see Section \ref{SIsec1:sec}) to a $\theta$-solvent chain conformation with $\nu \approx 0.5$. Fig. (\ref{Fig4:SIfig}) therefore shows selected results from non-equilibrium simulations for $\varepsilon = 0.1~k_{\mathrm{B}}T$ (good solvent conditions) and $1.0~k_{\mathrm{B}}T$ (poor solvent conditions). Fig. (\ref{Fig4:SIfig}a) and (\ref{Fig4:SIfig}b) show that the more attractive $\varepsilon = 1.0~k_{\mathrm{B}}T$ leads to a similar diffusion $D$ compared to the more repulsive good solvent counterpart, but the emerging reaction rate constants $k$ are increased in the diffusion-controlled limit. Higher values of $\varepsilon$ cause faster reactions due to the more attractive nature. \\
The size scaling in non-equilibrium for the good solvent case with $\varepsilon = 0.1~k_{\mathrm{B}}T$ is comparable to the equilibrium as demonstrated by the Fig. (\ref{Fig4:SIfig}c), (\ref{Fig4:SIfig}d), and (\ref{Fig4:SIfig}f). There, the choice of the reaction propensity does not affect the size of the resulting chain structures. however, for poor solvent conditions, a drastic shift towards higher values of $\nu$ is observed with an increasing propensity. This manifest in the snapshots (Fig. (\ref{Fig4:SIfig}f)), where less compact coil structures appear for fast reactions with $\lambda\tau_{\mathrm{B}} = 10^{5}$. The higher $\nu$ from non-equilibrium chain polymerization in poor solvent conditions compared to their equilibrium equivalents originates from the search for additional monomers and required relaxation time after addition of more monomer to the chain. The density profiles of Fig. (\ref{Fig4:SIfig}e) demonstrate the excluded volume for both values of $\varepsilon$, and also shows the more attractive nature of $\varepsilon = 1.0~k_{\mathrm{B}}T$, which is pronounced for slow reactions, where free monomer can accumulated around the active site between two reaction events. In the case of fast reactions, the two different solvent conditions lead to similar density profiles relatively close to the Smoluchowski limit for ideal, diffusion-controlled reactions. \\   

\section{Density effects on growing chains at the absence of pairwise interactions}
\label{SIsec5:sec}
\begin{figure}[!htbp]
\begin{center}
\includegraphics[width=8.6cm]{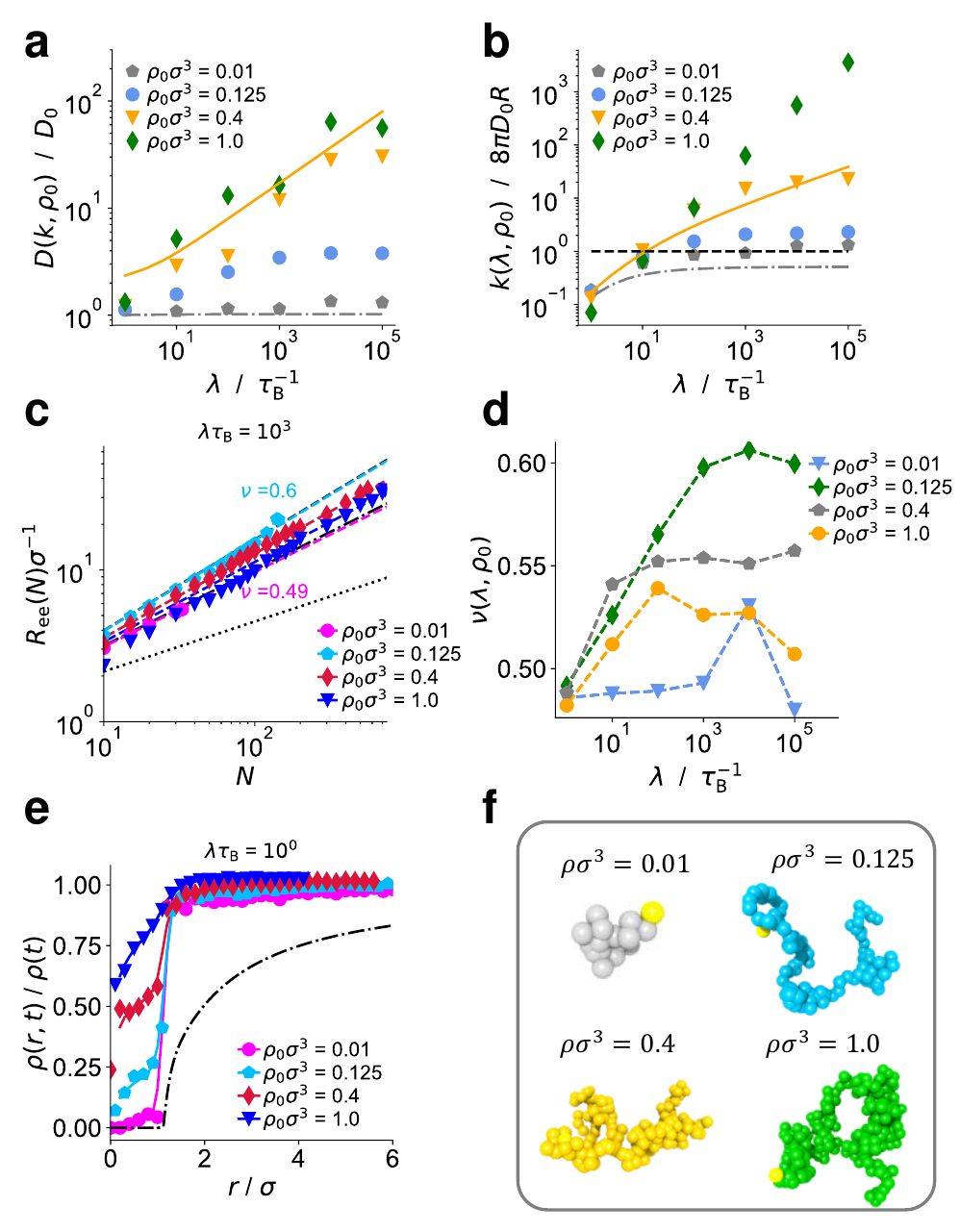}
\caption{\textbf{Growth without interactions at different initial densities.} (a) Diffusion coefficients $D$ (symbols) for different reaction rates $k$ and initial densities $\rho_{0}$. The orange line and the dashdotted gray line represent the fastest and slowest theoretical limit for $\rho_{0}\sigma^{3} = 1.0$ and $0.01$, respectively. (b) Simulated reaction rate constants $k$ (symbols) for different reaction propensities $\lambda$ and initial densities $\rho_{0}$. The orange line and the dashdotted gray line represents the fastest and slowest theoretical limit for $\rho_{0}\sigma^{3} = 1.0$ and $0.01$, respectively, and the dashed black line the fastest Smoluchowksi reaction rate $k_{\mathrm{S}} = 8 \pi D_{0}R$.  (c) End-to-end distance at different densities for an intermediate reaction frequency $\lambda\tau_{\mathrm{B}} = 10^{3}$. Symbols show selected simulated data points, colored dashed lines represent corresponding fits with $R_{\mathrm{ee}}(N) \propto N^{\nu}$. Black dashdotted, dashed, and dotted lines represent theoretical curves for $\nu = 3/5$ (good solvent), $\nu = 1/2$ ($\theta$-solvent), and $\nu = 1/3$ (poor solvent). (d) Size scaling exponents $\nu$ for different propensities $\lambda$. (e) Normalized densities profiles $\rho(r,t)/\rho(t)$ around the active site for different reaction conditions compared with the theoretical profile in the diffusion-controlled ('Smoluchowski') limit $\rho(r,t) = \rho(t)(1-R/r)$ (black, dashdotted line)~\cite{Smoluchowski1918, Dibak2019}. (f) snapshots of fast growing chains ($\lambda\tau_{\mathrm{B}} = 10^{5}$) with yellow-colored active sites.}
\label{Fig5:SIfig} 
\end{center}
\end{figure}

Fig. (\ref{Fig4:fig}b) in the main article contains simulated data points describing the scaling behavior for initial densities ranging from $\rho_{0}\sigma^{3} = 0.01$ to $1.0$. Both, the lower and upper simulated limit for the density show reduced scaling exponents $\nu$ for the different reaction propensities $\lambda$. For a low density $\rho_{0}\sigma^{3} = 0.01$, this is not surprising, since the number of reactions events and jumps is low and reactions kinetics in the fast limit are comparable to the Smoluchowski rate constant for bimolecular reactions of two moving particles (Fig. (\ref{Fig5:SIfig}a) and (\ref{Fig5:SIfig}b)). The simulated data points are found close to our suggested theoretical slow limit for the mutual diffusion based on Equation~\ref{DiffFast:eq}. High densities $\rho_{0}\sigma^{3} = 0.4$ and $1.0$ lead to fast reactions, and since the later one is above the critical percolation density $\rho_{\mathrm{crit}}\sigma^{3} \approx 0.43$ (Eq.~\ref{Rate:perc}), extremely high reaction rates $k$ are reported, which are only limited by the propensity $\lambda$ and the technical constraint of allowing only two bond formations per reaction time-step. It should be noted that only $\rho_{0}\sigma^{3} = 0.4$ is more or less still covered by theoretical data using Eq.~\ref{DiffFast:eq}, but no meaningful theory is accessible beyond $\rho_{\mathrm{crit}}$ due to divergence of the numerical solutions. \\
Figs. (\ref{Fig5:SIfig}c), (\ref{Fig5:SIfig}d), and (\ref{Fig5:SIfig}f) demonstrate the different size scaling $R_{\mathrm{ee}} \propto N^{\nu}$ behavior dependent on the density. Low densities of $\rho_{0}\sigma^{3} = 0.01$ and thus slow reactions allow a relaxation of chains between reaction events and structures with $\nu \approx 1/2$ representing $\theta$-conditions for all reaction propensities $\lambda$ studied. As shown in the article, densities between $\rho_{0} = 0.125$ and $0.3$ lead to enhanced scaling exponents close to the good solvent limit $\nu \approx 3/5$. Higher densities $\rho_{0}\sigma^{3} > 0.3$ don't lead to a further increase of $\nu$, but rather show again a decreasing tendency back towards random walk and $\theta$-solvent behavior with increasing density. \\
We explain this non-monotonic behavior of the scaling exponent $\nu$ for an increasing density with an increased number of monomers in each reactive volume around an active site (Fig. (\ref{Fig5:SIfig}e)) and the increased reaction rate (Fig. (\ref{Fig5:SIfig}b)). At high densities close to and above the critical density $\rho_{\mathrm{crit}}\sigma \approx 0.43$, potentially monomer-depleted zones are filled up quickly and orientation memory $\tau_{\mathrm{rot}}$ represented by a few chain elements is negligible with respect to the growth rate. Form a macroscopic perspective, the chain growth performs a random walk through the dense monomer melt. The enhanced size scaling exponent is only observed for densities below a critical percolation threshold $\rho_{\mathrm{crit}}$.
\newpage

\bibliography{210701Chaingrowth_Sprints}

\subsection*{Author contributions}
M. B. and J. D. conceived the research, M. B. performed the numerical simulations and the corresponding analysis of data. All authors participated in the analysis and the interpretation of the results, as wall as the writing of the manuscript.

\end{document}